\documentstyle[sprocl]{article}

\input{psfig}

\def\Journal#1#2#3#4{{#1} {\bf #2}, #3 (#4)}

\def\etal{{\it et al.}}

\def\AA{\em A.\& A.}
\def\APJ{\em ApJ.}
\def\APP{\em Astropart. Phys.}
\def\CPC{\em Computer Physics Commun.}
\def\IMP{\em Int. J. Mod. Phys.}
\def\JPG{\em J. Phys. G: Nucl. Part. Phys.}
\def\JPg{\em J. Phys. G}
\def\JPL{\em JETPhys. Lett.}
\def\MPA{{\em Mod. Phys.} A}
\def\NAT{\em Nature}

\def\NPB{{\em Nucl. Phys.} B}

\def\PRL{\em Phys. Rev. Lett.}
\def\PRD{{\em Phys. Rev.} D}
\def\PRE{\em Phys. Rep.}

\def\be{\begin{equation}}
\def\ee{\end{equation}}
\def\bea{\begin{eqnarray}}
\def\eea{\end{eqnarray}}

\newcommand{\lfig}[2]{\psfig{file=#1,width=\figwidth,angle=#2}}

\newdimen\captwidth
\newdimen\figwidth
\newcommand{\figfr}[1]{
\def\fighfrac{#1}
\captwidth=\linewidth
\figwidth=\fighfrac\linewidth
\advance\captwidth by -\figwidth
\advance\captwidth by -4truemm
}

\begin {document}

\title {Can Dark Matter be Ultra Heavy Particles ?}

\author{H. Ziaeepour}
\address{European Southern Observatory(ESO),\\Schwarzchild strasse, 2\\85748, Garching b. M\"{u}nchen, Germany}

\maketitle\abstracts {The detection of High Energy Cosmic Rays (HECR) with 
energies around and 
beyond GZK expected cutoff has introduced the idea of existence of a 
decaying Ultra Heavy Dark Matter (UHDM). If this type of particles make a 
substantial part of the CDM, their decay can have important implications 
for evolution of the large structures and high energy backgrounds. Here 
we present preliminary results of numerical solution of Boltzmann 
equations in presence of a decaying CDM. The evolution 
of baryons and photons energy are calculated. We find that in a homogeneous
universe, UHDM can not have a large contribution to the CDM. However, this 
hypothesis can probably survive if one takes into account the clumpiness of 
the matter in the Universe.}

\section {Motivations and Evidences for a Very Heavy Dark Matter}

One of the recently raised and yet unexplained issues in Particle Physics is 
the observation of Ultra High Energy Cosmic Rays (UHECRs)(For a review see
~\cite{crrev}). The source of these 
particles can not be in a distance larger than $30-50 Mpc$, otherwise, 
interaction with CMBR and IR photons considerably reduces the energy of the 
primary particles~\cite{cmbir}. This phenomena should produce a steep cutoff 
(GZK cutoff) in the spectrum of CRs at energies around $\sim 10^{18} eV$ to 
$10^{20} eV$. Observation shows a local 
minimum in the spectrum around these 
energies, but unexpectedly it rises again at higher energies. In fact the 
spectrum is compatible with two distinct class of sources, one for energies 
less than $\sim 10^{18} eV$, and the other for higher energies.
It is very difficult to detect the composition of UHECR primaries. It seems 
however that the data is more compatible with proton or neutrino like 
primaries~\cite{compo}.
\\
No candidate source has been observed in the direction of UHECRs in a distance 
permitted by GZK cutoff. N. Hayashida \etal~\cite{18ev} reports a small excess 
of $\sim 4\%$ for UHECRs with energies $\sim 10^{18} eV$ in the direction of 
the Galactic Center and Cygnus region. However, it is shown~\cite{nocorr} 
that there is not any correlation between observed events and any of 
individual galaxies in the $50 Mpc$ radius for events with energies 
$> 10^{18} eV$.\\
Standard production mechanism of UHECRs is the 
acceleratation of charged particles in the shock waves of AGNs, SNs, or in-falling 
gas in rich clusters. The main condition for these objects to be able to 
produce the observed spectrum of UHECRs is the existence of relatively strong 
magnetic field with large coherence length. The secondary acceleration of 
charged particles by 
remnant of old SNs can accelerate protons up to $\sim 10^{15} eV$ and iron 
to $\sim 10^{18} eV$~\cite{src1}. Models for proton acceleration in AGNs can 
produce UHECRs up to energies 
$\sim 10^{20} eV$ ~\cite{accagn} with a marginal maximum energy of 
$4 \times 10^{21} eV$~\cite{maxacc}. They also predict a change of composition 
from iron to proton. However, the uncertainty on jet density, coherence length 
of the magnetic field in hot spots and large distance to the candidate 
sources incompatible with energy loss due to interaction with CMB, make these 
sources marginal candidates.\\
Gamma Ray Bursts (GBR) also have been proposed as the candidate source for 
UHECRs~\cite{grb}. If UHECRs are produced in a relativistic shock wave, one 
expect to observe after-glows of this process in lower energies (like GBRs). 
Attempts to find an excess of $TeV$ range $\gamma-$Rays in the direction of 
observed UHECR events have failed~\cite{burst}. One more argument against an 
acceleration source for UHECRs is that one expects to observe a bunch of 
UHECRs at least for the most nearby sources as it is the case for GRB. No 
such event has been yet observed.\\
Another possible class of sources for these events is the decay of a very 
heavy elementary particle with a GUT-scale mass~\cite{xpart}. These particles 
can be either long life very heavy particles, or short life particles produced 
from the decay of topological defects. In the latter case, the heavy 
particles decay in their turn to ordinary particles and make the observed 
UHECRs. 
There is a number of mechanisms which can produce long life ultra heavy 
particles in the early universe. At the end of an inflationary epoch, 
the parametric resonance of inflaton field oscillation can produce very 
heavy particles with masses larger than inflaton mass itself~\cite{reso}. 
In super-symmetric models where SUSY breaking in the hidden sector is 
communicated to the observable sector by gauge fields, the existence of a 
global symmetry can lead to stability or long life of lightest particle. The 
lightest messenger gauge field get also a large mass and under some symmetry 
conditions can have a long life~\cite{susy}. Vacuum fluctuation of a background field coupled to the gravitation 
also can produce very heavy particles with ${\Omega}_X \approx 1$ for a range of inflaton mass and self coupling~\cite{grapro}. This makes these particles a 
very good candidates for Dark Matter (DM). The unitarity constraint on the mass of DM particles~\cite {uni} can be 
turned over if they have never been thermalized. The condition of a long life 
time for a very heavy particle can be fulfilled if it has some discrete gauge 
symmetry~\cite{dissy}. The possibility that topological defects be the source of UHECRs has been also 
studied extensively~\cite {defect}. Some years ago, there was a lot of hopes 
and attempts to prove that topological 
defects can be the seed of LSS and CMB fluctuations (see ~\cite{toprev} for a 
review). However, the results of 
comparison between simulations and observed power spectrum of LSS and CMB 
anisotropy practically rules out this possibility~\cite{againstdef}.\\
By contrast, as mentioned above, some models can 
produce a CDM totally composed of ultra heavy particles~\cite {grapro}. 
Their decay can have 
important implication on the evolution of LSS. Remnants 
contribute to the total amount of baryonic matter in the universe mostly as a 
non-thermalized hot component. Another important implication is the production 
of large amount of high energy backgrounds. This can be used as a test 
for verification of this 
hypothesis. In this work, we calculate the opacity of 
the universe for remnants with highest energies and the evolution of their 
energy spectrum. We compare expected flux if CDM is UHDM, with observations.
\section {Model}
We assume that the $X$ particles are neutral bosons with a mass 
$m_X = 10^{24} eV$ and a life time $\tau = 1.2 \time 10^{10} yr$. 
If the decay of these particles looks like the decay of $Z^{\circ}$, theoretical and 
experimental arguments show that leptonic and hadronic channels have a 
branching ratio of $\sim 1/3-2/3$. Other channels have a negligible 
contribution. As the dominant channel is the hadronic one and the leptonic 
channel makes very soon a cascade, here we only 
consider this mode of decay. It is very likely that $X$ particles don't decay 
directly to known particles. This leads to softening of the energy spectrum.
For simulation of hadronization, we consider the hadronization of two gluon 
jets. Experimental observation as well as MLLA (Modified Leading Logarithm 
Approximation)~\cite {hadr}, LPHD (Local Parton-Hadron Duality)(~\cite {lphd} 
and references therein) and string hadronization model~\cite {strhadr} 
obtain a softer spectrum with higher multiplicity for gluon jets than for 
quark jets. This is probably more convenient for the decay of $X$ particles.

\begin{figure}[t]
\figfr{0.60}
\hbox to\linewidth{
\lfig{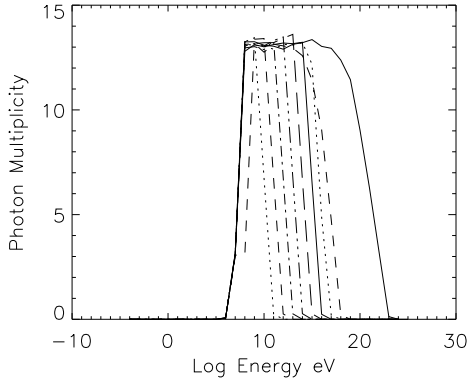}{0}
\hfill\parbox[b]{\captwidth}{
\caption[short caption]{Photon multiplicity in hadronization of a pair of gluon jets with CM energies from $10^{12} eV$ to $10^{24} eV$.}
\label{fig:decay} 
}}
\end{figure}

We use PYTHIA program~\cite{pythia} for jet hadronization. This program, 
like many other available ones, can not properly simulate ultra high energy events, 
not only because we don't know the exact physics at $10^{16} GeV$ scale, but also 
because of programming limits. For this reason, we had to extrapolate simulation 
results from lower energies. Fig.\ref {fig:decay} shows as an example, 
the photon multiplicity for hadronization of a pair of gluon jets. 
At energies larger than $10^{12} eV$, the multiplicity per $log (E) \sim y$ 
(rapidity) is roughly constant. The same behavior exists for other species. 
It is the result of KNO scaling predicted by MLLA. We use this 
property to extrapolate the multiplicity spectrum to invariant CM mass of 
$10^{24} eV$. It should be mentioned that at high energies, one expects that 
KNO scaling be violated. The fact that in Fig.1 we don't see this effect shows 
that at high energies, the results of simulation programs can be uncertain. We think 
that various parameterizations of hadronic spectrum also suffer from the same 
problem, simply because they have been fitted to low energy models or 
observations. In this respect, simulations are probably more close to 
reality because they take into account more non-perturbative effects.
In the simulation, we assume that all particles except $e^{\pm}, p^{\pm}, \nu, 
\bar {\nu} \& \gamma$ will decay. The mean energy contribution of the stable 
species is as the following: $13\%$ $e^{\pm}$, $25\%$ $p^{\pm}$, $36\%$ $\nu \& 
\bar {\nu}$ \& $25\%$ $\gamma$. More than $99\%$ of the 
total energy belongs to the particles with energies 
higher than $10^{20} eV$. If baryogenesis happens in GUT scale, we expect 
an excess of baryons over their anti-particles. We assume a positive 
baryonic number and a null leptonic number. This results to an excess of 
$e^-$ and $p^+$ over $e^+$ and $p^-$ (to keep the universe electrically 
neutral), and $\bar{\nu}$ over ${\nu}$. For simplifying our simulation, we 
assume that all $\nu$' s are $\nu_e$ and massless. It is interesting to note 
that $\nu$' s have the largest contribution to the total decay energy.\\
The decay of a very heavy particle has two implications: Energy dissipation of 
the remnant in one hand heats the ordinary (visible) matter in the Universe and 
on the other hand it contributes itself to the yield of the visible matter. 
To calculate the spectrum of the remnant, one has to solve the complete 
Einstein-Boltzmann equations. Even in the case of linearized equations, the 
presence of interactions makes this system of equations non-linear and coupled. 
Specially, in the case of UHECR energy dissipation, the problem is highly 
stiff in the sens that one has to treat $29$ orders of magnitude in energy, 
from CMB energy to the mass of $X$ particles. The numerical solution of these 
equations is under test~\cite {houri}. Here we discuss the results of a more 
simplified model.
We consider a homogeneous universe and we neglect the 
production of high energy particles due to interaction of remnants with 
backgrounds. The reason behind this assumption, apart from simplifying the 
model, is that we are only interested in comparing the predicted flux for 
very energetic particles $E > 10^{17} eV$ with observation, as we know that 
lower energies have other sources than presume UHDM and thus, they are not 
directly signals of these particles.
\begin{figure}[t]
\figfr{0.60}
\hbox to\linewidth{
\lfig{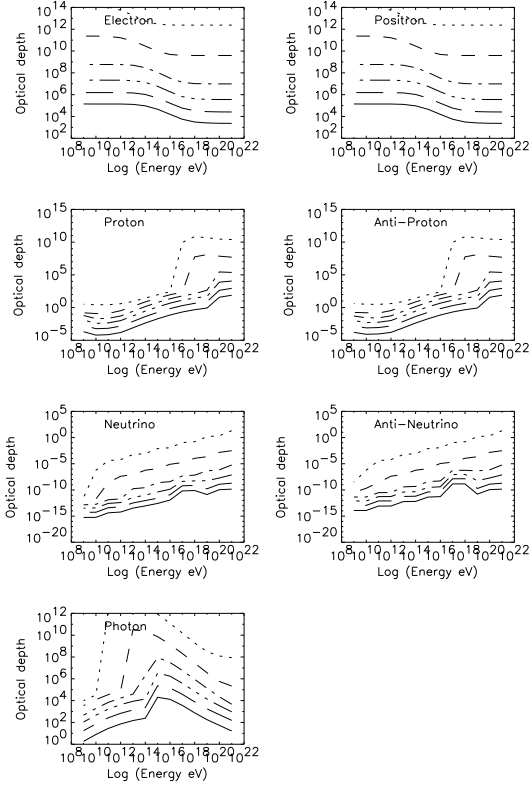}{0}
\hfill\parbox[b]{\captwidth}{
\caption[short caption]{Optical depth versus energy at different red-shift. 
Energies are particle energy at $z = 0$. Solid $z = 0.02$, long dashes 
$z = 0.1$, dash 3-dot $z = 0.5$, dash dot $z = 2$, dashed $z = 10$, dot 
$z = 50$.}
\label{fig:opte} 
}}
\end{figure}
We assume that particle energy completely dissipates and makes a 
high energy background. For $\gamma$ and neutrinos, we assume that this 
background has a Gaussian distribution. For massive particles i.e $e^{\pm}$ 
and $p^{\pm}$, we assume that they have a fixed mean temperature. 
In one hand this simplification is motivated by the results of 
S. Lee in\cite {crpro} where propagation of UHECRs has been studied for a flat 
homogeneous universe. He considers also the effect of production of high 
energy particles from interaction of remnants with backgrounds and show 
that there is an increase of photon background at energies lower than 
$\sim 10^{14} eV$. On the other hand, there are 
numerous evidence of existence of a hot IGM, ICM (intra-cluster medium). 
\begin{figure}[t]
\figfr{0.60}
\hbox to\linewidth{
\lfig{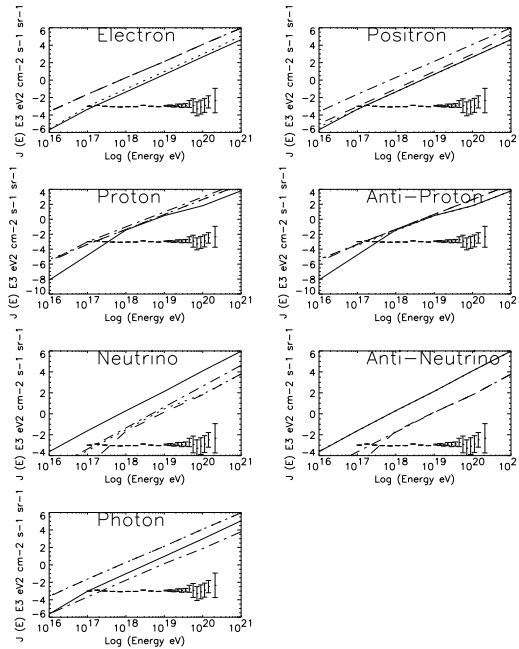}{0}
\hfill\parbox[b]{\captwidth}{
\caption[short caption]{Energy Flux at $z = 0$ with 4 different $\gamma$ 
backgrounds: Solid and dotted $E_{bg} = 10^{10}eV$, dashed and dash 
$E_{bg} = 10^6eV$, and respectively with and without radio background. Dots present observed flux. Only highest energies should be compared with observation.}
\label{fig:e2dens} 
}}
\end{figure}
Backgrounds else than remnants are BBN baryons, primordial neutrinos and 
CMB and star light. We use models developed by Salamon \& 
Stecker~\cite{opti} and by Vernet \& Valls-Gabaud~\cite{joel} respectively 
for IR and visible and for UV backgrounds.
We assume $\Omega_m = 0.3$ and $\Omega_{\Lambda} = 0.7$.\\
As expected, the optical depth depends on the temperature assumed for the 
remnant background. Neutrinos temperature is assumed to be $10^{12} eV$. The 
reason for this choice is that at lower energies their cross-section is 
very small. Remnant photons have significant role on the opacity of high 
energy particles. We studied following temperatures:
$10^4 eV$, $10^5 eV$, $10^6 eV$ and $10^{10} eV$. Each of these temperatures 
has its maximum effect for a range of remnant energies. 
Fig.\ref{fig:opte} shows as an example, the optical depth for $E^{\gamma}_{bg} = 10^6 eV$.
Reducing background 
energy increases the optical depth because the number of photons 
will increase.\\
Fig.\ref {fig:e2dens} shows the density of stable particles at $z = 0$. 
Irrespective of background photon energies, the expected flux of UHECRs at 
$z = 0$ from the model is much 
larger than observed flux. This result definitively rule out the existence of 
large contribution of UHDM in the CDM or its life time must be at least $4$ 
order of magnitude larger than the age of the Universe (see also~\cite {sarkar}). However, we believe 
that our model 
is too simplified. The Universe is far from having a uniform distribution of 
matter. Dark Matter as well as baryonic matter is clumpy. The opacity for 
remnant is proportional to the density of the background. 
Fig.\ref {fig:opclust} shows the optical depth of a clump with a mean 
over-density 
of $\frac {\delta \rho}{\rho} = 200$ and a radius of $1 Mpc$. Apart from 
$\nu$s, all other high energy species dissipate their energies in the clump. 
This results to a large reduction of expected flux on the 
Earth and can probably survive the UHDM model. More results about the 
effect of clumpiness on the UHECRs will be reported elsewhere.\\
I am grateful of J. Vernet and D. Valls-Gabaud for providing me with the 
results of their UV background model.
\begin{figure}[t]
\figfr{0.60}
\hbox to\linewidth{
\lfig{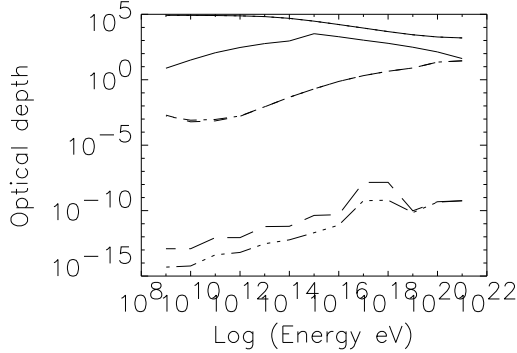}{0}
\hfill\parbox[b]{\captwidth}{
\caption[short caption]{Optical depth of a clump with radius = $1 Mpc$ and 
$\frac {\delta \rho}{\rho} = 200$ for $e^{\pm}$ and $\gamma$ solid line, 
$p^{\pm}$ dash dot, $\nu$ dash 3-dot and $\bar {\nu}$ long dashes.}
\label{fig:opclust} 
}}
\end{figure}

\end{document}